\documentclass[fleqn,usenatbib,useAMS]{mnras}
\usepackage{graphicx}   
\usepackage{amsmath}    
\usepackage{amssymb}    
\usepackage{multicol}        
\usepackage{bm}         
\usepackage{pdflscape}  
\usepackage[T1]{fontenc}
\usepackage{ae,aecompl}
\usepackage{newtxtext,newtxmath}

\def\aap{A\&A}
\def\apjl{ApJ}
\def\apj{ApJ}
\def\apjs{ApJS}
\def\aj{AJ}
\def\mnras{MNRAS}
\def\nat{Nature}
\def\pasp{PASP}

\def\araa{ARAA}

\title[The Most Massive Pulsating White Dwarf]{WD J004917.14$-$252556.81, the
Most Massive Pulsating White Dwarf}
\author[Kilic et al.]
{Mukremin Kilic$^1$, Alejandro H. C\'orsico$^{2,3}$, Adam G. Moss$^1$, Gracyn Jewett$^1$, \newauthor Francisco C. De Ger\'onimo$^{4,5}$, Leandro G. Althaus$^{2,3}$\\
$^1$Homer L. Dodge Department of Physics and Astronomy, University of Oklahoma, 440 W. Brooks St., Norman, OK 73019, USA\\
$^2$Grupo de Evoluci\'on Estelar y Pulsaciones, Facultad de Ciencias Astron\'omicas y Geof\'{\i}sicas, Universidad Nacional de La Plata,Paseo del Bosque s/n,\\ 1900 La Plata, Argentina\\
$^3$IALP - CONICET, La Plata, Argentina\\
$^4$Instituto de Astrof\'isica, Pontificia Universidad Cat\'olica de Chile, Av. Vicu\~na Mackenna 4860, 7820436 Macul, Santiago, Chile \\ 
$^5$Millennium Institute of Astrophysics, Nuncio Monse\~nor Sotero Sanz 100, Of. 104, Providencia, Santiago, Chile}

\date{\ \ Submitted \today \vspace{-0.5cm}}
\pubyear{2023}

\begin{document}
\label{firstpage}
\pagerange{\pageref{firstpage}--\pageref{lastpage}}
\maketitle

\begin{abstract}

We present APO and Gemini time-series photometry of WD J004917.14$-$252556.81, an ultramassive DA white dwarf
with $T_{\rm eff} = 13020$ K and $\log{g} = 9.34$. We detect variability at two significant frequencies,
making J0049$-$2525 the most massive pulsating white dwarf currently known with $M_\star=1.31~M_{\odot}$ (for a CO core)
or $1.26~M_{\odot}$ (for an ONe core). J0049$-$2525 does not display any of the signatures of binary mergers, there is
no evidence of magnetism, large tangential velocity, or rapid rotation. Hence, it likely formed through single star evolution
and is likely to have an ONe core. Evolutionary models indicate that its interior is $\gtrsim99$\% crystallized.
Asteroseismology offers an unprecedented opportunity to probe its interior structure. However, the relatively few pulsation modes
detected limit our ability to obtain robust seismic solutions. Instead, we provide several representative solutions that could explain
the observed properties of this star. Extensive follow-up time-series photometry of this unique target has the potential to
discover a significant number of additional pulsation modes that would help overcome the degeneracies in the asteroseismic
fits, and enable us to probe the interior of an $\approx1.3~M_{\odot}$ crystallized white dwarf.

\end{abstract}


\section{Introduction}

The majority of stars that evolve in isolation end up as CO core white dwarfs \citep{fontaine01}. Mass transfer
in a binary can cut the evolution of a star short, leading up to the formation of low-mass white dwarfs with
He cores and masses below about $0.45~M_{\odot}$. On the opposite end of the mass spectrum, 
off-center carbon ignition in a degenerate CO core with $M\geq1.06~M_{\odot}$ should lead
to the formation of ONe core white dwarfs \citep{murai68}. However, binary mergers may also form CO core white dwarfs in the same
mass range \citep[e.g.,][but see \citealt{schwab21}]{althaus21}. Observationally, it is nearly impossible to
constrain the core composition of a white dwarf. 

\citet{kepler16b} reported the discovery of a white dwarf with an atmosphere dominated by O, Ne, and Mg
with abundance ratios of O/Ne = 25 and O/Mg = 55. The lack of any H or He lines in the spectrum of this star, along
with its ONeMg atmosphere, make it an excellent candidate for an ONe core white dwarf.
Asteroseismology is the only method that can probe the interior structure and core composition
\citep{fontaine08,winget08,giammichele18,corsico19b}, though with few exceptions, the majority of the pulsating white dwarfs
known have $M<1.06~M_{\odot}$, and therefore are expected to have CO cores.

There are three pulsating white dwarfs with mass estimates above $M_\star>1.1~M_{\odot}$ as reported in the literature. 
BPM 37093 \citep{kanaan92} is a $T_{\rm eff} = 11620 \pm 189$ K and $M_\star=1.13 \pm 0.14~M_{\odot}$ \citep{bedard17}
DAV white dwarf with pulsation periods of 512-635 s \citep{metcalfe04}. GD 518 shows pulsations with periods
of 425-595 s \citep{hermes13}, and has $T_{\rm eff}=12,030 \pm 210$ K and
$M_\star = 1.23 \pm 0.03~M_{\odot}$ based on the 1D atmosphere models and a CO core \citep{gianninas11}. 
However, a recent analysis based on Gaia parallax indicates a lower mass, with the best-fitting parameters
of $T_{\rm eff}=11422 \pm 106$ K and $M_\star = 1.114 \pm 0.006~M_{\odot}$ \citep{kilic20}. 
SDSS J084021.23+522217.4 is another DAV with pulsations periods of 173-797 s \citep{curd17} and
has a spectroscopic mass estimate above  $1.1~M_{\odot}$, but the photometric method and Gaia parallax
indicate a mass below $1~M_{\odot}$ in the Montreal White Dwarf Database \citep{dufour17}.

\begin{figure*}
\centering
\includegraphics[width=3.2in, clip=true, trim=0.2in 0.3in 0.7in 0in]{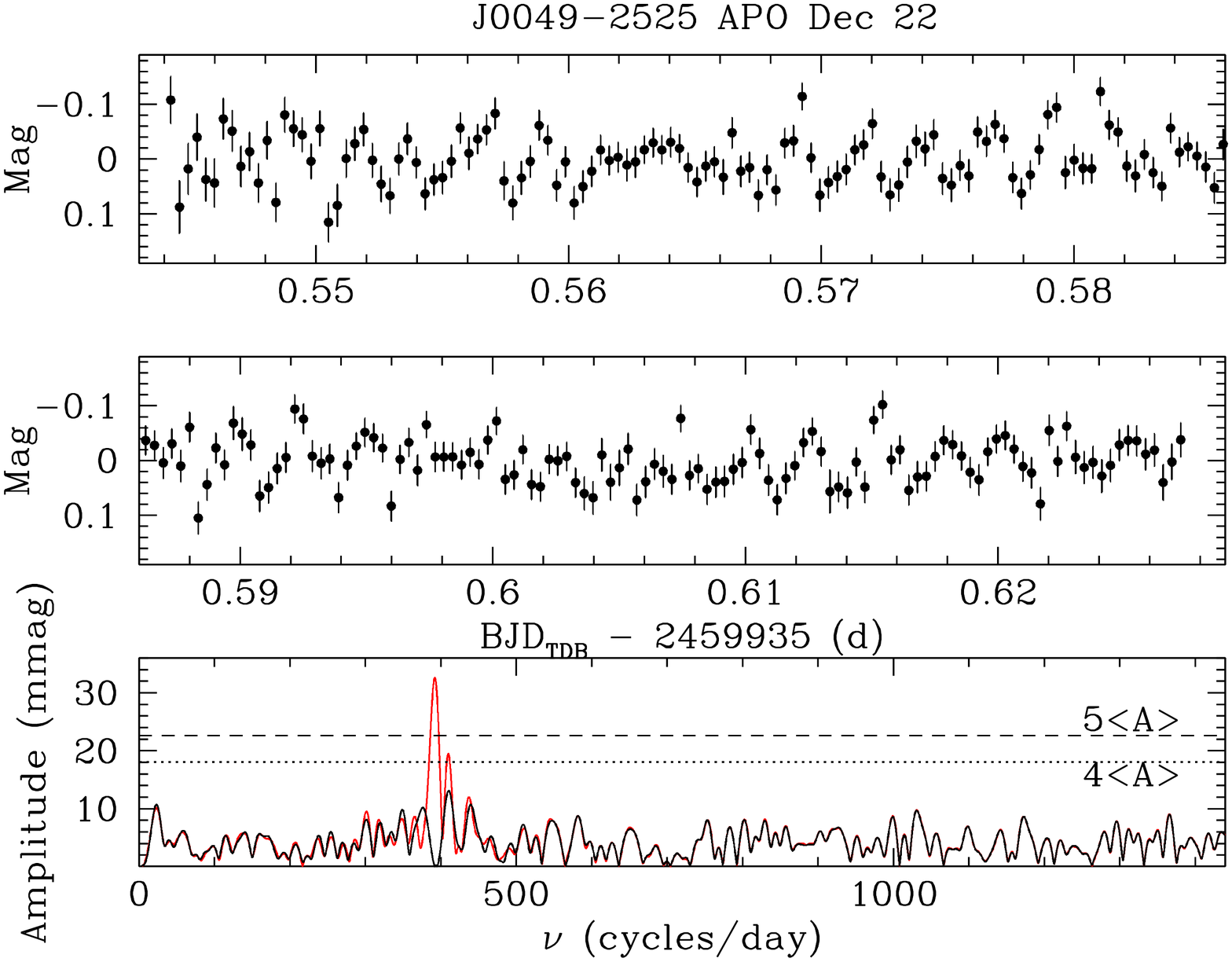}
\includegraphics[width=3.2in, clip=true, trim=0.2in 0.3in 0.7in 0in]{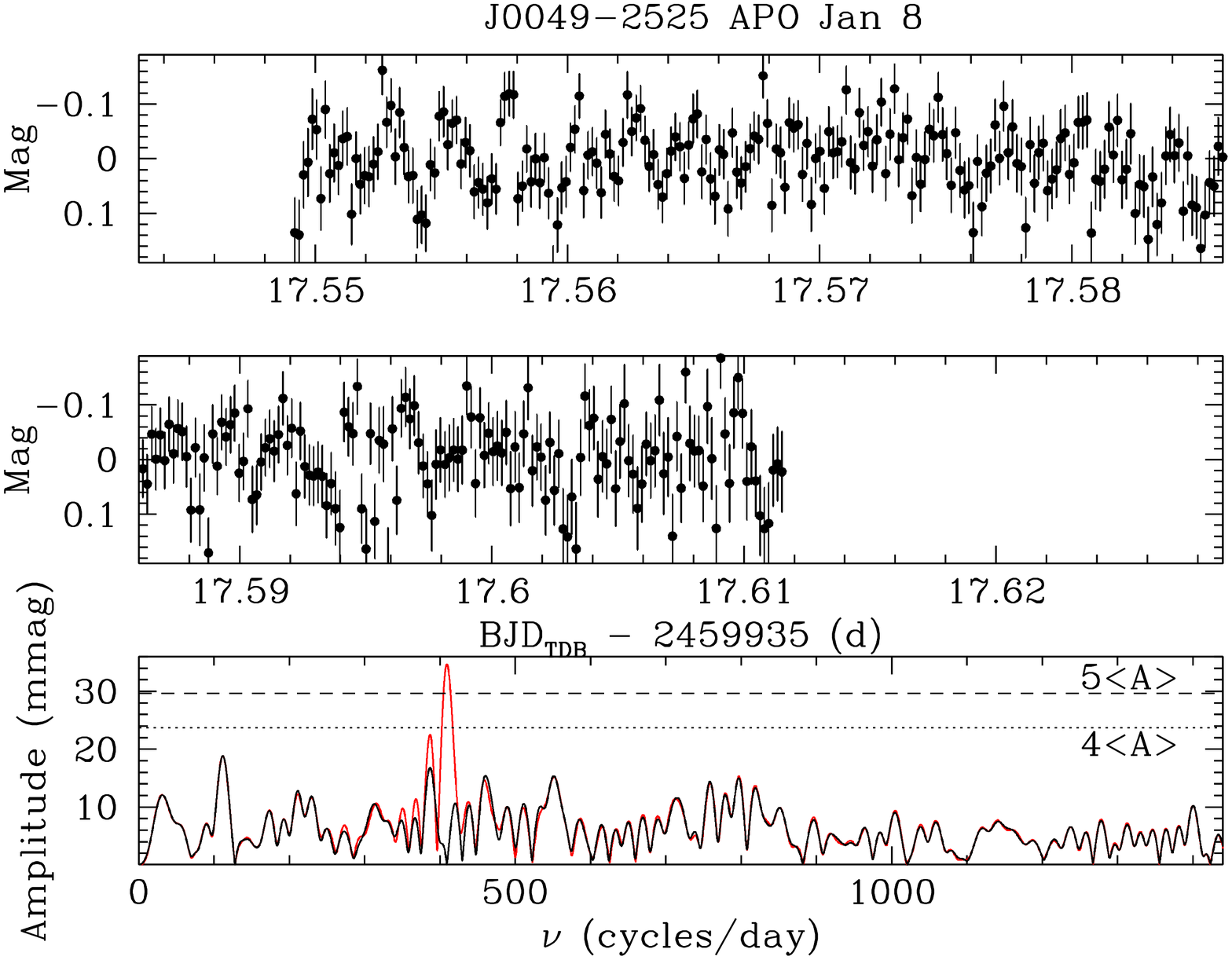}
\caption{APO time-series photometry of J0049$-$2525 on two separate nights (top panels).
The bottom panels show the Fourier transform of each light curve before (red) and after
(black) prewhitening of the dominant frequency. The dotted and dashed lines
mark the 4 and 5$\langle {\rm A}\rangle$ level, where $\langle {\rm A}\rangle$
is the average amplitude in the Fourier transform.}
\label{figapo}
\end{figure*}

\begin{figure*}
\centering
\includegraphics[width=3.2in, clip=true, trim=0.2in 0.3in 0.7in 0in]{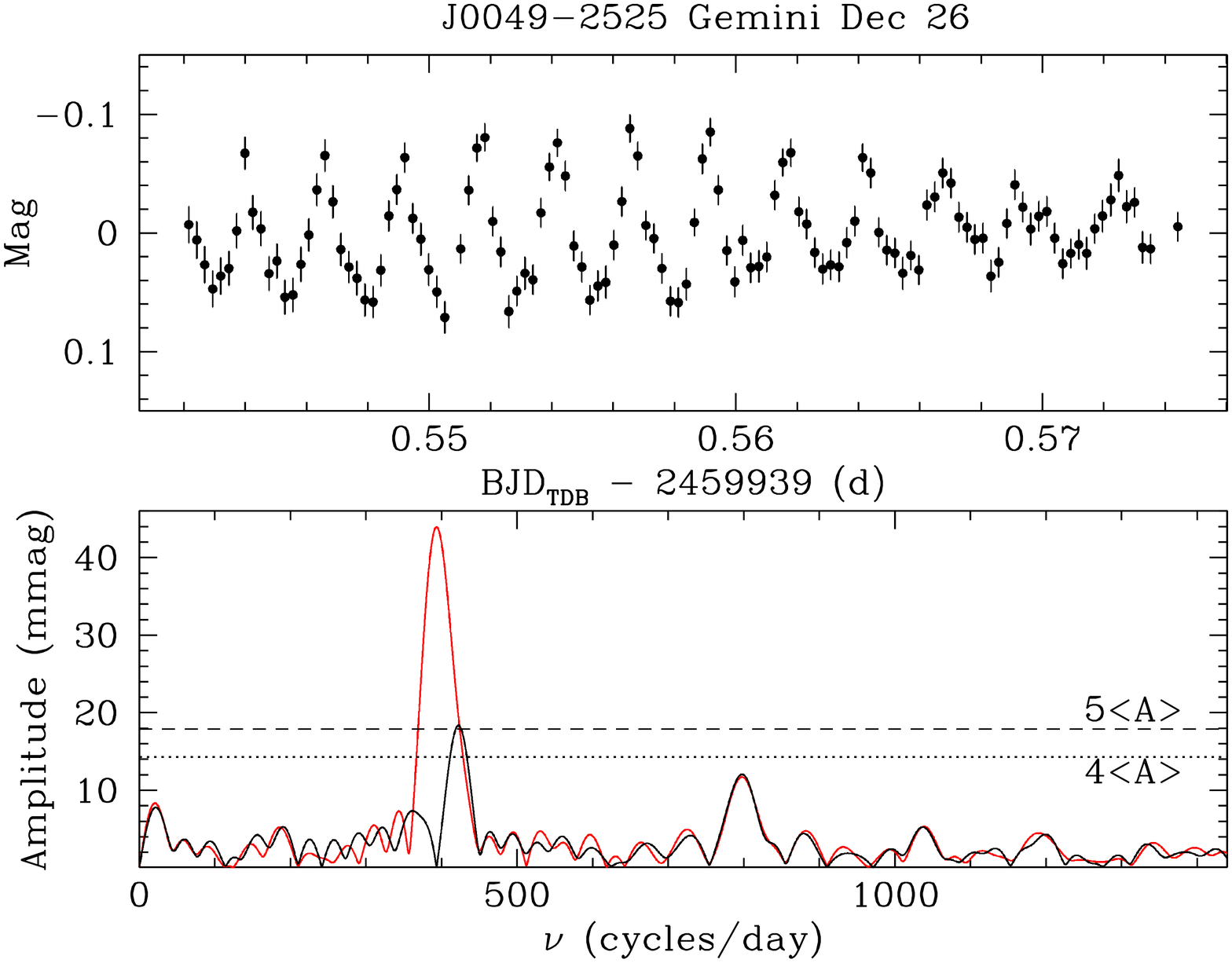}
\includegraphics[width=3.2in, clip=true, trim=0.2in 0.3in 0.7in 0in]{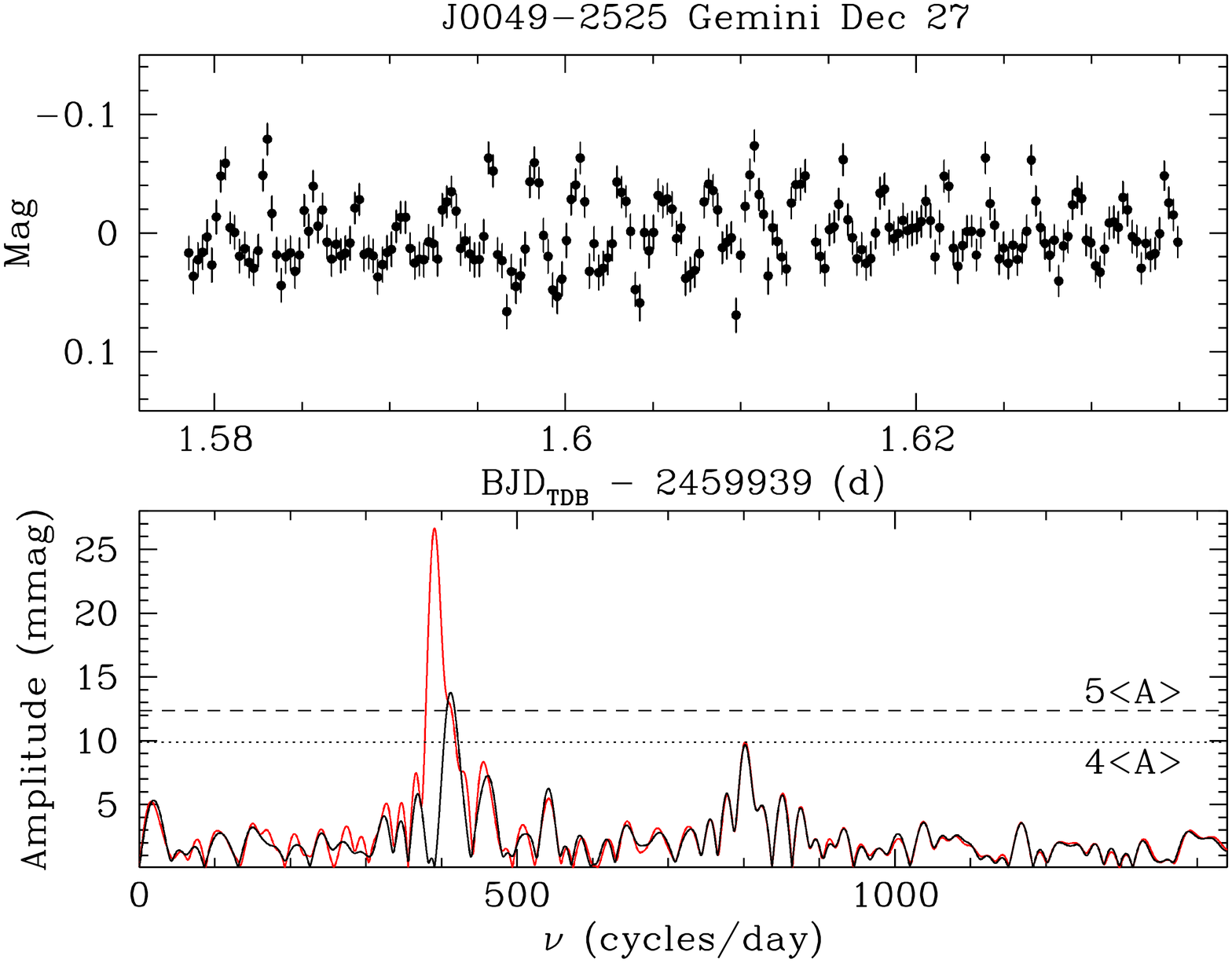}
\caption{Gemini time-series photometry of J0049$-$2525 obtained over two consecutive nights. The panels and labels
are the same as in Figure \ref{figapo}. The observations span 47 (left) and 81 min (right), respectively.}
\label{figgem}
\end{figure*}

\citet{rowan19} and \citet{vincent20} identified three additional pulsating white dwarf candidates with a mass 
$\geq1.05~M_{\odot}$. However, none of these three objects have follow-up spectroscopy available in the literature,
and they all show variations at a single dominant period of 330, 357, and 809 s, respectively. Rapidly rotating
white dwarfs are known to show photometric variations at similar periods \citep{pshirkov20,caiazzo21}, and they may impersonate ZZ Ceti white
dwarfs \citep{kilic21b}. Hence, further spectroscopy and time-series photometry are necessary to confirm that
these three targets are indeed DA white dwarfs with multi-periodic photometric variations due to pulsations.

Recently, \citet{kilic21} presented an analysis of the ultramassive ($M_\star\geq1.3~M_{\odot}$) white dwarf candidates
in the Montreal White Dwarf Database 100 pc sample, and identified 25 white dwarfs with masses
up to $1.35~M_{\odot}$, assuming CO core composition. \citet{kilic23} presented spectroscopic and photometric
follow-up observations of this sample to constrain the merger fraction among this population. In this process, they also identified
an exciting ZZ Ceti candidate. Table \ref{tabpar} presents the physical parameters of WD J004917.14$-$252556.81,
which is a DA white dwarf with $T_{\rm eff} = 13020 \pm 460$ K and $\log{g}= 9.341 \pm 0.036$. These parameters
make it by far the best ZZ Ceti candidate in the 100 pc sample of ultramassive white dwarfs. However, J0049$-$2525 has
no prior time-series photometry available.

\begin{table}
\centering
\caption{Physical Parameters of J004917.14$-$252556.81}
\begin{tabular}{cc}
\hline
Parameter & Value \\
\hline
Spectral Type & DA \\
$T_{\rm eff}$ (K) & 13020 $\pm$ 460 \\
$\log{g}$ & $9.341 \pm 0.036$ \\
Distance (pc) & $99.7^{+2.9}_{-2.7}$ \\
Mass, ONe core ($M_{\odot}$) & 1.263 $\pm$ 0.011 \\
Cooling age, ONe core (Gyr) & 1.94 $\pm$ 0.08 \\
Mass, CO core ($M_{\odot}$) & 1.312 $\pm$ 0.010 \\
Cooling age, CO core (Gyr) & 1.72 $\pm$ 0.09 \\
\hline
\end{tabular}
\label{tabpar}
\end{table}

Here we present APO and Gemini time-series photometry of J0049$-$2525. We detect multi-periodic photometric
variability, making J0049$-$2525 the most massive pulsating white dwarf currently known. We present the details
of our observations in Section 2, the pulsation spectrum of J0049$-$2525 in Section 3, and results from our representative
asteroseismic models in Section 4. We conclude in Section 5.

\section{High-cadence Photometry}

We acquired high speed photometry of J0049$-$2525 on UT 2022 Dec 22 using the APO 3.5m telescope with the Agile
frame transfer camera \citep{mukadam11} and the BG40 filter. We obtained back-to-back exposures of 30 s over 2 hours.
We binned the CCD by $2\times2$, which resulted in a plate scale of $0.258\arcsec$ pixel$^{-1}$. Given the southern declination
of this target, our observations were obtained under $2.5\arcsec$ seeing and an airmass of 2. In addition, due to the $2.2\arcmin$
field of view of Agile, there was only one reference star available. Even with these setbacks, we detected significant photometric
variations in J0049$-$2525, which prompted us to ask for the Director's Discretionary Time on Gemini South.

We obtained Gemini GMOS time-series photometry of J0049$-$2525 on UT 2022 Dec 26 and 27 as part of the program
GS-2022B-DD-107. We obtained 121 and $215\times7$ s back-to-back exposures on Dec 26 and 27, respectively. We used the
SDSS-$g$ filter, and binned the chip by 4$\times$4. This resulted in a plate scale of $0.32\arcsec$ pixel$^{-1}$ and a 15.7 s overhead,
resulting in a cadence of 22.7 s. We used four reference stars brighter than J0049$-$2525 for relative photometry.

We obtained additional APO photometry of J0049$-$2525 on UT 2023 Jan 8 using the same setup as above. We obtained 
$360\times15$ s back to back exposures with Agile.

\section{Pulsations in J0049$-$2525}

Figure \ref{figapo} shows the APO light curves of J0049$-$2525 from UT 2022 Dec 22 and 2023 Jan 8 and their Fourier transforms (bottom panels).
The black lines show the original spectrum (shown in red) pre-whitened by the dominant peak. The data from the first night shows two frequencies,
near 390 and 410 cycles per day, above the 4$\langle {\rm A}\rangle$ level, where $\langle {\rm A}\rangle$ is the average amplitude in the Fourier
transform. The dominant frequency has $\approx30$ millimag amplitude, and is detected at high significance. However,
the pre-whitening of the Fourier transform by the dominant frequency lowers the amplitude of the secondary peak to below
the 4$\langle {\rm A}\rangle$ level.   

Interestingly, these two frequencies seem to be persistent in the second night of APO data (see the right panels in Figure \ref{figapo}).
Here the dominant frequency has actually switched. The second frequency near 410 cycles per day is detected above the 5$\langle {\rm A}\rangle$ level with
an amplitude of 35 millimag, while the first frequency near 390 cycles per day is present at below the 4$\langle {\rm A}\rangle$ level. Since both
frequencies are persistent on two nights of observations from APO, and both show amplitude variations, these data clearly indicate that J0049$-$2525
shows multi-periodic variability due to pulsations. Further support comes from the Gemini data.

Figure \ref{figgem} shows the Gemini light curves of J0049$-$2525 from UT 2022 Dec 26 and 27 along with their Fourier transforms.
Even though Gemini observations have a shorter baseline, the data quality is significantly better. These light curves show peak-to-peak
variations that vary over time. For example, the peak-to-peak variation is 0.15 mag at 2459939.55 d in the top left panel, whereas it is
0.07 mag at 2459939.57 d. Similar variations are also visible in the top right panel. Hence, there is clearly more than one frequency at play. 
Fourier transforms of each light curve shows that there are three significant frequencies detected, F1 at $\approx$390, F2 at $\approx410$, and the
combination frequency F1+F2 at $\approx800$ cycles per day. F2 is resolved better in the second night's data given its longer baseline.
In addition, both F1 and F2 are detected above the 5$\langle {\rm A}\rangle$ level in both nights of Gemini observations. The amplitudes of the two modes appear to
change from night to night, similar to what is seen in the APO data, and also similar to the ultramassive white dwarf GD 518 \citep{hermes13}. 

Tables \ref{tabapo} and \ref{tabgem} present the results from the frequency analysis of the APO and Gemini data, respectively.
We report the results from both a least square analysis and a Monte Carlo approach where we replace each photometric measurement
$m$ with $m + g~\delta m$, where $\delta m$ is the error in magnitude and $g$ is a Gaussian deviate with zero mean and unit variance.
For each of the 1000 sets of modified light curves for each night, we calculate the Fourier transform using the {\tt Period04} package
in batch mode \citep{period04}, and we take the range that encompasses 68\% of the probability distribution function as the $1\sigma$
uncertainties in frequency and amplitude. The results from the least squares and the Monte Carlo analysis agree within the errors, but we adopt
the results from the latter as it provides a better estimate of the error distribution.

\begin{table}
\caption{Multi-mode frequency solutions based on the APO data.}
\begin{tabular}{cccccc}
\hline
Date& ID & Method&  Frequency & Period & Amplitude \\
        &   &  & (cycles d$^{-1}$) &  (s) & (millimag)  \\
\hline
2022 Dec 22 & F1  & LSQ & $391.64 \pm 0.72$ & $220.61 \pm 0.40$ & $30.3 \pm 3.3$ \\
2023 Jan 8   &  F2  & LSQ & $409.14 \pm 1.19$ & $211.18 \pm 0.61$ & $34.6 \pm 4.7$ \\
\hline
2022 Dec 22 & F1  & MC & $391.94_{-0.90}^{+0.75}$ & $220.44_{-0.42}^{+0.50}$ & $32.2_{-3.3}^{+2.8}$ \\
2023 Jan 8   &  F2  & MC & $409.13_{-1.23}^{+1.29}$ & $211.18_{-0.66}^{+0.63}$ & $35.0_{-4.3}^{+4.2}$ \\ 
\hline
\end{tabular}
\label{tabapo}
\end{table}

\begin{table}
\caption{Multi-mode frequency solutions based on the Gemini data.}
\begin{tabular}{cccccc}
\hline
Date& ID & Method&  Frequency & Period & Amplitude \\
        &     &  & (cycles d$^{-1}$) &  (s) & (millimag)  \\
\hline
2022 Dec 26 & F1 & LSQ & $388.34 \pm 0.67$ & $222.48 \pm 0.38$ & $41.9 \pm 1.6$ \\
\dots              & F2 & LSQ & $418.36 \pm 1.20$ & $206.52 \pm 0.59$ & $23.1 \pm 1.6$ \\
\dots              & F1+F2 & LSQ & $796.06 \pm 2.33$ & $108.53 \pm 0.32$ & $12.0 \pm 1.6$ \\
2022 Dec 27 & F1 & LSQ & $390.22 \pm 0.57$ & $221.42 \pm 0.32$ & $27.7 \pm 1.6$ \\
\dots              & F2 & LSQ & $412.16 \pm 1.09$ & $209.63 \pm 0.55$ & $14.5 \pm 1.6$ \\
\dots              & F1+F2 & LSQ & $802.02 \pm 1.63$ & $107.73 \pm 0.22$ & $ 9.6 \pm 1.6$\\
\hline
2022 Dec 26 & F1 & MC & $388.33_{-1.05}^{+1.00}$ & $222.49_{-0.57}^{+0.61}$ & $41.8_{-1.7}^{+2.0}$\\
\dots             & F2 & MC & $418.47_{-1.78}^{+1.44}$ & $206.47_{-0.71}^{+0.88}$ & $23.3_{-1.7}^{+2.0}$\\
\dots       & F1+F2 & MC & $795.93_{-4.55}^{+3.49}$ & $108.55_{-0.47}^{+0.62}$ & $12.2_{-1.6}^{+1.7}$\\
2022 Dec 27 & F1 & MC & $390.13_{-0.53}^{+0.55}$ & $221.46_{-0.31}^{+0.30}$ & $27.9_{-1.3}^{+1.4}$\\
\dots              & F2 & MC & $412.38_{-1.29}^{+1.41}$ & $209.52_{-0.71}^{+0.66}$ & $14.9_{-1.4}^{+1.5}$\\
\dots              & F1+F2 & MC & $801.26_{-\dots}^{+1.88}$ & $107.83_{-0.25}^{+\dots}$ & $10.2_{-1.1}^{+1.1}$\\
\hline
\end{tabular}
\label{tabgem}
\end{table}

The APO data reveal two dominant modes at $220.44_{-0.42}^{+0.50}$ and $211.18_{-0.66}^{+0.63}$ s with 32-35 millimag amplitudes,
whereas the Gemini data show the same modes at $222.49_{-0.57}^{+0.61}$ and $206.47_{-0.71}^{+0.88}$ s on 2022 Dec 26, and
$221.46_{-0.31}^{+0.30}$ and $209.52_{-0.71}^{+0.66}$ s on Dec 27.  It is possible that the observed modes in J0049$-$2525 are unstable,
and the observed frequency variations in the Gemini data are real. However, due to the relatively short baseline (47 min) of the observations
on Dec 26, the dominant peak in the Fourier transform is relatively broad, and our measurements of the frequencies are likely impacted by the observing
window. Hence, the frequency measurements from Dec 27 are superior, and they are also consistent with the APO data within $2\sigma$. 

Combining APO and Gemini results, the weighted mean frequencies are F1 = $221.36 \pm 0.23$ s and F2 = $209.38 \pm 0.40$ s. 
The observed multi-periodic variations in J0049$-$2525 over four different nights confirm it as the most massive pulsating white dwarf currently
known. J0049$-$2525 is significantly more massive than the previous record holders BPM 37093 and GD 518. In addition, the latter objects
show variations at the 1-4 mma level \citep{metcalfe04,hermes13}, whereas J0049$-$2525 has pulsations at the 30 millimag level,
making it much easier to detect its variability.

\section{Discussion}

\subsection{Asteroseismology of Ultramassive White Dwarfs}

Ultramassive white dwarfs are expected to have a significant portion of their cores in a crystalline state at the effective temperatures
near the ZZ Ceti instability strip. For J0049$-$2525, the crystallized fraction is expected to be $\gtrsim99$\%.

Asteroseismology can provide insight into the interior structure of ZZ Ceti white dwarfs, although periods ($\Pi$) and period spacings ($\Delta \Pi$)
vary with mass, effective temperature, H envelope mass, and the crystallized mass fraction. In other words, there are four parameters to fit, but the crystallized
mass fraction is also a function of the stellar mass and temperature and therefore is not independent. Crystallization (i) shrinks the resonant cavity, which
results in a growth in the period spacing, additional to that due to cooling only; (ii) engulfs the entire region of the core, and these regions and their chemical
compositions become impossible to probe with pulsations since the $g$-modes cannot penetrate the crystallized matter. Due to crystallization, the
mode trapping properties and the resulting period-spacing features are largely simplified because only the He/H transition remains active. The results
of (i) and (ii) are a $\Delta \Pi$ distribution with much less features even for short periods (low order modes), and mean period spacing much higher
than what would be without crystallization \citep[see][]{montgomery99,corsico05}. 

The fact that g-mode eigenfunctions cannot penetrate the solid regions of the core makes it virtually impossible to discriminate the deep core chemical
composition (CO versus ONe) with pulsations in the most ultramassive DAV white dwarfs. However, for less massive ones, e.g. with $M_\star\sim1.1~M_{\odot}$,
features of the outermost parts of the core can be probed by means of pulsations, which may be enough to distinguish between CO and ONe cores
\citep[see Fig. 2 and 3 from][]{degeronimo19}.

\citet{corsico19} performed asteroseismological analysis of several pulsating ultramassive WDs, among them, BPM 37093, the richest pulsator known.
Using the 8 modes identified by \citet{metcalfe04}, the authors were able to perform period spacing analysis and period-to-period fits. On the other
hand, for GD 518, which exhibits just three modes \citep{hermes13}, only period-to-period fits were performed.
For BPM 37093, the mean period spacing of $\sim17$ s corresponds to $\ell = 2~g$-modes.
Due to the degeneracy of the dependence of the mean period spacing on stellar mass, effective temperature, and the H envelope mass, \citet{corsico19}
could not independently infer the mass of the star, but they could restrict possible solutions by comparing the observed period spacing with the
average theoretical period spacings. They found a best-fit model by considering the individual pulsation periods. The best-fitting asteroseismological model for BPM 37093 has
$T_{\rm eff} = 11650$ K, $M_\star = 1.16~M_\odot$, $\log{(M_H/M_\star)} = -6$, and a crystallized mass fraction of 92\%.

The case of ultra-massive DBVs, if they exist, would be more encouraging, because in this case, crystallization in the DBV instability strip
($T_{\rm eff} \sim 24000$-30000 K) is not capable of hiding the chemical composition of the core completely \citep{corsico21}. 

\subsection{Representative Asteroseismological Models for J0049$-$2525}

With only two significant modes detected, it is impossible to perform a detailed asteroseismological analysis of J0049$-$2525. 
For instance, in order to put constraints on the mass of the H envelope, it is necessary to have many periods available in such a way
that one can construct period-spacing diagrams ($\Delta \Pi$ versus $\Pi$). The quantity $\Delta \Pi$ is very sensitive to $M_H$. In particular,
thick H envelopes give rise to short trapping cycles (intervals between $\Delta \Pi$ minima) and low amplitudes of trapping (magnitude of $\Delta \Pi$
minima), while thin H envelopes result in long trapping cycles and very deep minima of $\Delta \Pi$ \citep{brassard92}.
Because we cannot create a period-spacing diagram for J0049$-$2525 based on only two frequencies, we have to resort to the individual period fits to find
representative seismological models.

\begin{figure*}
\centering
\includegraphics[width=5in, clip=true, trim=0.1in 1.2in 1.9in 1.2in]{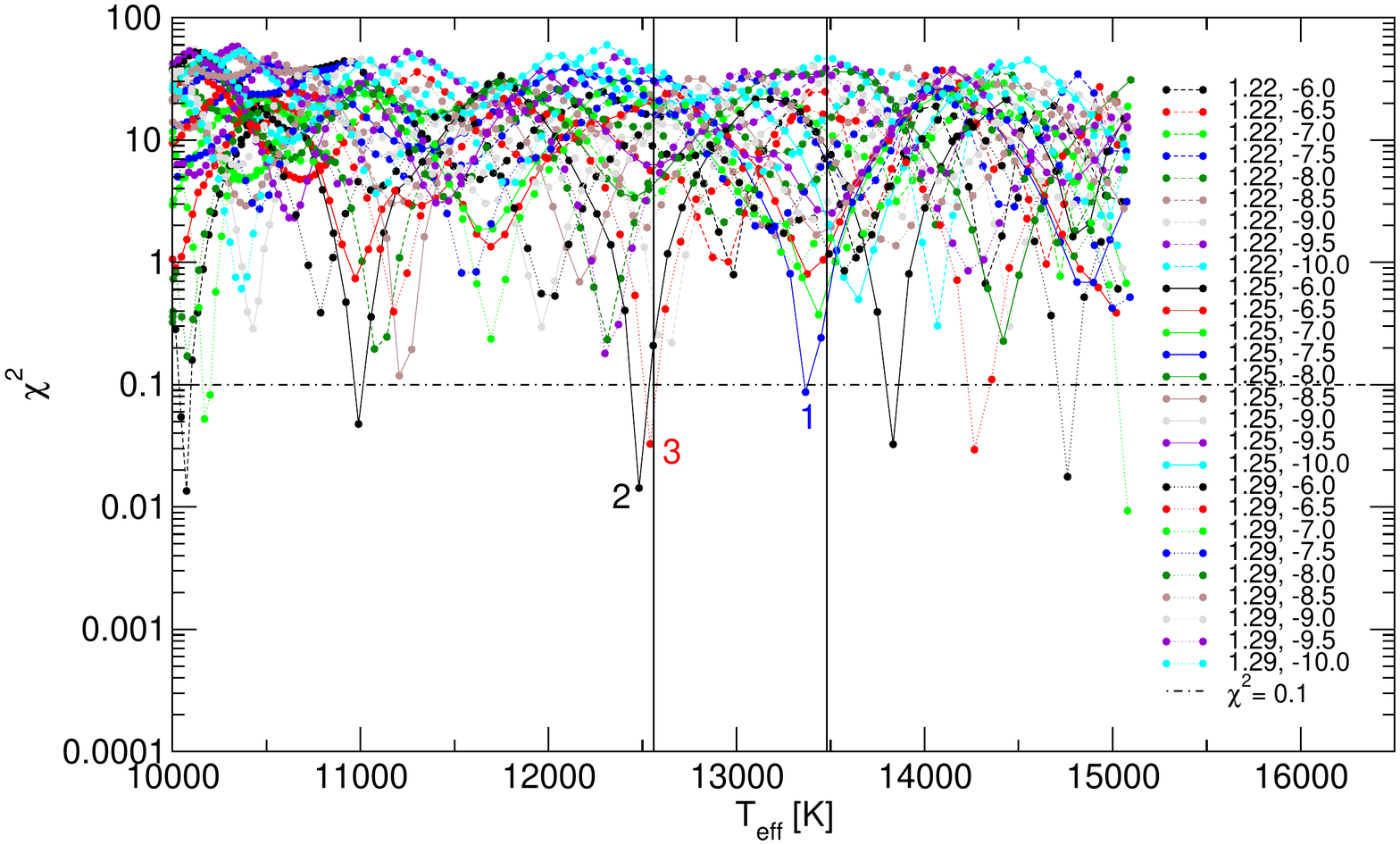}
\includegraphics[width=5in, clip=true, trim=0.1in 1.2in 1.9in 1.2in]{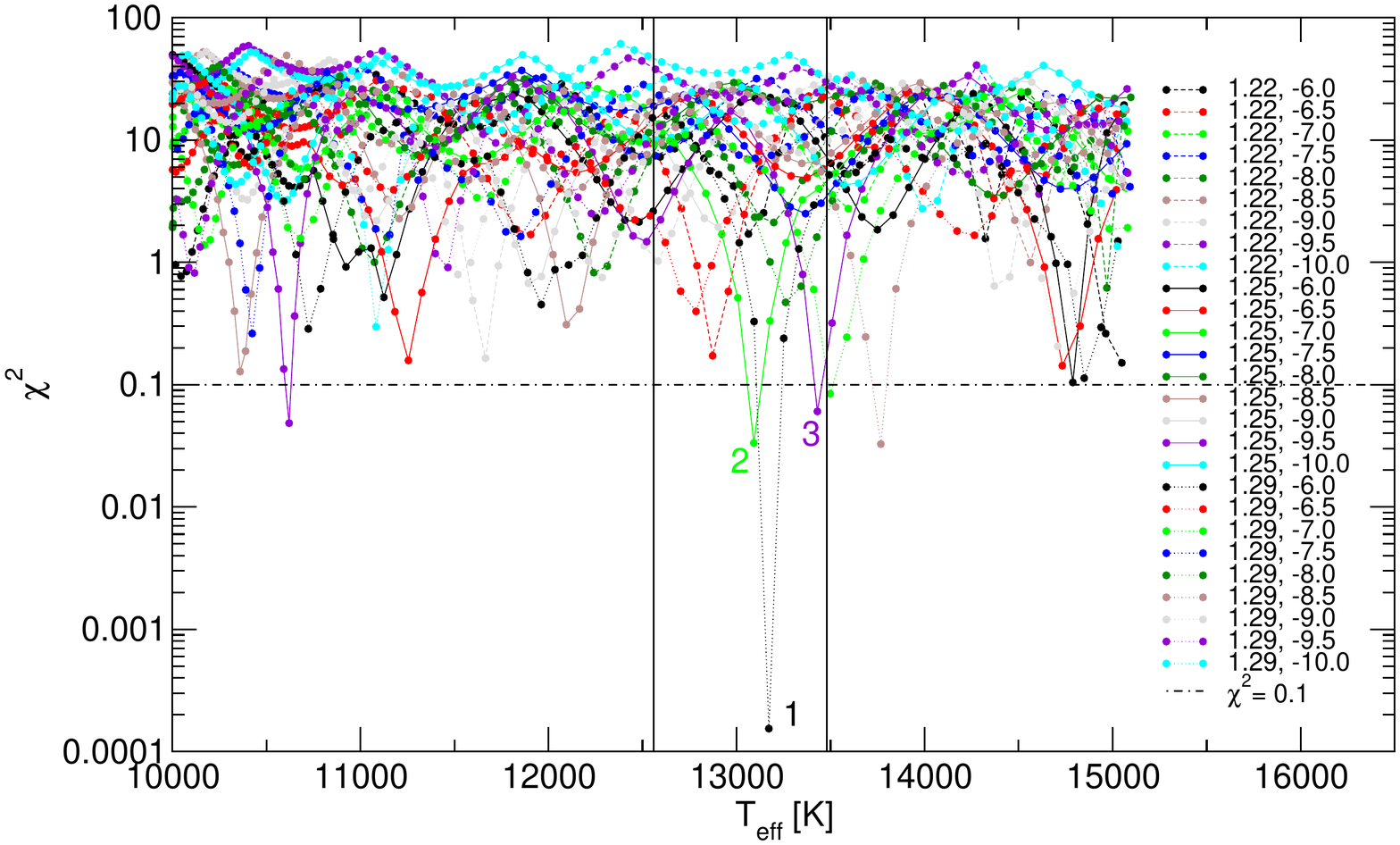}
\caption{The $\chi^2$ of the period fit as a function of the effective temperature for the ultramassive DA white dwarf
model sequences with $M_\star=1.22$ to $1.29~M_{\odot}$ and H envelopes ranging from $10^{-6}$ to $10^{-10}~M_{\star}$.
Vertical solid lines show the $\pm1\sigma$ range of the spectroscopic effective temperature for J0049$-$2525.
The top panel shows the fits to the weighted mean periods obtained from APO and Gemini data,
whereas the bottom panel shows the fits using the periods measured only from the APO data.
Three best-fitting models are labelled as 1, 2, and 3 in each panel, and their parameters are given in Table \ref{tabmod}.}
\label{figale}
\end{figure*}

We use an expanded grid of evolutionary sequences of ultramassive white dwarfs with ONe cores and stellar masses $M_\star = 1.10$,
1.13, 1.16, 1.19, 1.22, 1.25, $1.29~M_{\odot}$ and H layer masses between  $10^{-6}$ and $10^{-10} M_\star$ (De Ger\'onimo et al 2023, in prep.).
The evolutionary models were computed using the LPCODE \citep{althaus05}, taking into account Coulombian diffusion \citep{althaus20}. Further
details are provided in \citet{camisassa19} and \citet{corsico19}. We computed adiabatic pulsation periods of $\ell = 1, 2$ g-modes in the range 70-1500 s,
as is typically observed in ZZ Ceti stars, employing the {\tt LP-PUL} pulsation code \citep{corsico19}.
The asymptotic period spacing for ultramassive DA white dwarfs with masses between 1.10 and $1.29~M_\odot$ and effective
temperatures within the ZZ Ceti instability strip (13500-10500 K) varies between $\sim22$ s and $\sim34$ s for $\ell = 1$, and between
$\sim12$ and $\sim19$ s for $\ell = 2$.

\begin{table}
\centering
\caption{Representative asteroseismic models for J0049$-$2525. The top (bottom) three lines correspond to the models
labelled in the top (bottom) panel of Figure \ref{figale}.}
\begin{tabular}{ccccccc}
\hline
Model& Mass & $\log{g}$ &  $T_{\rm eff}$ & $\log{\frac{M_H}{M_\star}}$ & $\frac{M_{\rm crys}}{M_{\star}}$ & d \\
        &  ($M_\odot$) & (cm s$^{-2}$)   & (K) &  & (\%) & (pc)  \\
\hline
1 & 1.25 &  9.274 & 13367  & $-7.5$ & 98.8 & $111.1 \pm 1.7$  \\
2 & 1.25 &  9.271 & 12482  & $-6.0$ & 98.7 & $104.4 \pm 1.6$\\
3 & 1.29 &  9.419 & 12541  & $-6.5$ & 99.6 & $91.0 \pm 1.4$ \\
\hline
1 & 1.29 &  9.414 & 13172 & $-6.0$ & 99.4 &  $95.2 \pm 1.5$ \\
2 & 1.25 &  9.276 & 13092 & $-7.0$ & 98.7 & $108.8 \pm 1.7$\\
3 & 1.25 &  9.278 & 13430 & $-9.5$ & 98.7 & $111.1 \pm 1.7$ \\
\hline
\end{tabular}
\label{tabmod}
\end{table}

With only two periods observed in J0049$-$2525, there are many possible models that reproduce both periods at the same time. 
The two periods are relatively close to each other (209 and 221 s). We adopted different scenarios for the mode identification:
(a) they are both $\ell= 1$ (or both $\ell= 2$) and consecutive modes, associated to a minimum in $\Delta \Pi$ due to mode
trapping. This is unlikely, but can be
checked by performing period-to-period fits assuming that both periods are $\ell= 1$ or 2; (b) one period is  $\ell= 1$ and the other one
is $\ell= 2$, which can also be checked in the period-to-period fits, leaving the mode identification as a free parameter;  and a more unlikely option
c) it is a rotational triplet ($\ell= 1$) where one of the components does not appear for some reason. In this case, (i) the components would be
$m=0$ and $m=+1$ or $m= -1$ (one of the latter is missing), or (ii) the components would be $m=+1$ and $m=-1$, with the $m=0$ component
missing. 

We carried out period fits considering (i) both periods as $\ell= 1$, (ii) both periods as $\ell= 2$, and (iii) one period is $\ell= 1$ and the other is $\ell= 2$,
without initially setting the $\ell$-value for each period. We adopted the weighted mean periods of 209.4 and 221.4 s. 
To evaluate the agreement between the theoretical and observed periods, we used the expression:

\begin{equation}
\chi ^2(M_{\star},M_H,T_{\rm eff})=\frac{1}{N} \sum_{i=1}^{N} \rm min[(\Pi_i^{O}-\Pi_k^{th})^2],  
\end{equation}

where $N$ is the number of detected modes, $\Pi_i^{\rm O}$ are the observed period and $\Pi_k^{\rm th}$ are the periods computed
theoretically ($k$ is the radial order). The best-fit model is selected by searching for the minimum value of $\chi^2$.  We did not consider
the theoretical density of modes here, because this is a preliminary analysis (due to having only 2 periods) and also because we do not have
a reliable recipe that takes into account the density of modes according to the harmonic degree and/or the stellar mass in a self consistent way.

In general, we obtained poor solutions when both modes are $\ell= 1$ or $\ell= 2$, and we can rule out these possibilities. On the other hand, a mixture
of $\ell= 1$ and $\ell= 2$ modes provide much better fits to the observed periods. Figure \ref{figale} shows the quality function $\chi^2$
as a function of the effective temperature of the star. 
Given the spectroscopic mass of the star ($1.26~M_\odot$ for an ONe core), here we only display models with $M_\star\geq1.22~M_\odot$.
For comparison, the top panel shows the fits to the weighted mean periods obtained from APO and Gemini data, whereas the bottom panel shows the fits
using the periods measured only from the APO data. In each case, there are many possible solutions with $\chi^2<0.1$, but there are a few that
fall within or near the $1\sigma$ range of the spectroscopic effective temperature of J0049$-$2525. We present the parameters of the 
three best-fit models in each case in Table \ref{tabmod}. These models indicate a crystallized mass fraction of 98.7 to 99.6\%.

A slight shift in the periods used in the model fits leads to very different results. The extreme sensitivity of the results to the precise values
of the periods used is due to the fact that there are only two periods observed. For example, we find an excellent fit to the periods derived
from the APO data only with a model that has $M_\star=1.29~M_\odot$, $T_{\rm eff} = 13172$ K, and $M_H= 10^{-6}~M_\star$. This is the model
labeled as 1 in the bottom panel of Figure \ref{figale}. The seismological distance of this model, $95.2 \pm 1.5$ pc, is also consistent with
the Gaia distance of $99.7^{+2.9}_{-2.7}$ pc \citep{bailer21} within the errors. Even though this is an excellent representative model of the star,
we caution that there are many other possible solutions, and we have to wait for the detection of additional periods in this object
to provide a more robust seismological solution. 

\section{Conclusions}
 
We report the discovery of multi-periodic variability in J0049$-$2525, a $\sim1.3~M_{\odot}$ ultramassive white dwarf within the 100 pc
sample. J0049$-$2525 does not display any of the signatures of binary mergers; there is no evidence of magnetism, rapid rotation, or
large tangential velocity ($V_{\rm tan} = 17$ km s$^{-1}$). Its spectrum is that of a typical DA white dwarf. Assuming that it formed through
single star evolution, it is likely to have an ONe core. The best-fit model to its observed spectral energy distribution places it within the boundaries
of the ZZ Ceti instability strip. 

\begin{figure}
\centering
\includegraphics[width=3.4in, clip=true, trim=0.3in 2in 0.6in 1.4in]{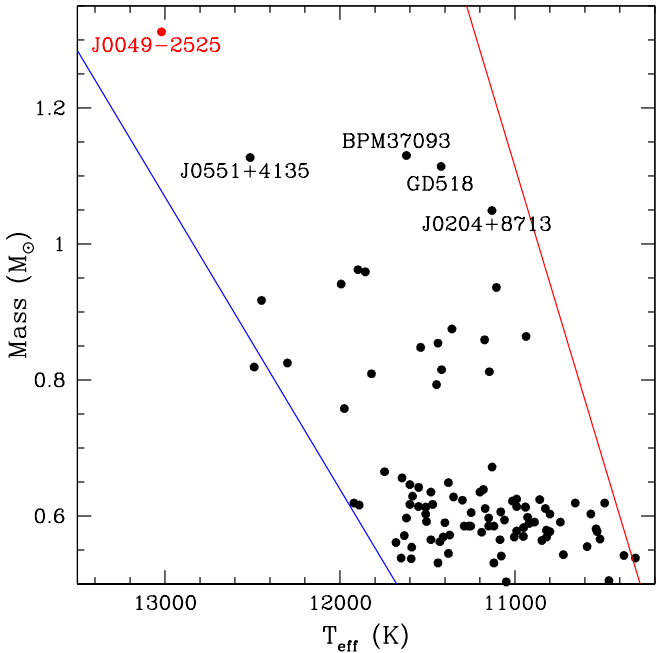}
\caption{Masses and effective temperatures for the previously known ZZ Ceti white dwarfs in the 100 pc
white dwarf sample from \citet{vincent20} assuming CO cores. The blue and red lines show the empirical boundaries of the
instability strip from the same work. Ultramassive ZZ Ceti white dwarfs with $M>1~M_{\odot}$ are labelled.}
\label{figstrip}
\end{figure}

We detected $\sim30$ millimag variations in J0049$-$2525 at two different frequencies over four different nights, confirming it
as the most massive pulsating DAV white dwarf currently known. Figure \ref{figstrip} shows the masses and effective temperatures
for the ZZ Ceti white dwarfs in the 100 pc white dwarf sample from \citet{vincent20} assuming CO cores. Ultramassive ZZ Ceti white dwarfs with
$M_\star>1~M_{\odot}$ are labelled. Note that two of these objects, J0551+4135 and J0204+8713, lack follow-up optical spectroscopy. Hence,
their classification as DAV white dwarfs requires confirmation. Another candidate that also needs follow-up spectroscopy for confirmation,
J212402.03$-$600100.0, was too far south to be included in the \citet{vincent20} study, and therefore not shown here.

J0049$-$2525 is significantly more massive than the other pulsating DAV white dwarfs known. Its mass is $1.312 \pm 0.010$ or
$1.263 \pm 0.011$ for a CO or ONe core, respectively \citep{kilic23}. Evolutionary models predict that its interior is $\gtrsim99$\% crystallized.
Asteroseismology offers a unique opportunity to probe the interior structure of crystallized massive white dwarfs like J0049$-$2525, provided
that a sufficient number of $g$-modes with consecutive radial order are detected. Because periods and their spacings vary with mass,
effective temperature, H envelope mass, and the crystallized mass fraction, and because J0049$-$2525 displays only two significant
modes in our APO and Gemini data, it is impossible to find a unique seismic solution. We provide several representative solutions that
can match the observed characteristics of this star. However, a significant number of additional pulsation modes would need to be detected
to overcome the degeneracies in the asteroseismic fits. We encourage extensive follow-up time-series photometry campaigns on this
unique target.

\section*{Acknowledgements}

This work is supported in part by the National Science Foundation under grants AST-1906379 and AST-2205736, the National Aeronautics
and Space Administration under grant 80NSSC22K0479, by AGENCIA through the Programa de Modernizaci\'on Tecnol\'ogica BID 1728/OC-AR,
and by the PIP 112-200801-00940 grant from CONICET.  
F.C.D.G. acknowledges financial support provided by ANID-FONDECYT grant 3200628.

The Apache Point Observatory 3.5-meter telescope is owned and operated by the Astrophysical Research Consortium.

Based on observations obtained at the international Gemini Observatory, a program of NSF's NOIRLab, which is managed by the Association of Universities for Research in Astronomy (AURA) under a cooperative agreement with the National Science Foundation on behalf of the Gemini Observatory partnership: the National Science Foundation (United States), National Research Council (Canada), Agencia Nacional de Investigaci\'{o}n y Desarrollo (Chile), Ministerio de Ciencia, Tecnolog\'{i}a e Innovaci\'{o}n (Argentina), Minist\'{e}rio da Ci\^{e}ncia, Tecnologia, Inova\c{c}\~{o}es e Comunica\c{c}\~{o}es (Brazil), and Korea Astronomy and Space Science Institute (Republic of Korea).

\section*{Data availability}

The data underlying this article are available in the Gemini Observatory Archive at
https://archive.gemini.edu, and can be accessed with the program number
GS-2022B-DD-107. The APO data that support the findings of this study are available
from the corresponding author upon reasonable request.


\begin{thebibliography}{}
\makeatletter
\relax
\def\mn@urlcharsother{\let\do\@makeother \do\$\do\&\do\#\do\^\do\_\do\%\do\~}
\def\mn@doi{\begingroup\mn@urlcharsother \@ifnextchar [ {\mn@doi@}
  {\mn@doi@[]}}
\def\mn@doi@[#1]#2{\def\@tempa{#1}\ifx\@tempa\@empty \href
  {http://dx.doi.org/#2} {doi:#2}\else \href {http://dx.doi.org/#2} {#1}\fi
  \endgroup}
\def\mn@eprint#1#2{\mn@eprint@#1:#2::\@nil}
\def\mn@eprint@arXiv#1{\href {http://arxiv.org/abs/#1} {{\tt arXiv:#1}}}
\def\mn@eprint@dblp#1{\href {http://dblp.uni-trier.de/rec/bibtex/#1.xml}
  {dblp:#1}}
\def\mn@eprint@#1:#2:#3:#4\@nil{\def\@tempa {#1}\def\@tempb {#2}\def\@tempc
  {#3}\ifx \@tempc \@empty \let \@tempc \@tempb \let \@tempb \@tempa \fi \ifx
  \@tempb \@empty \def\@tempb {arXiv}\fi \@ifundefined
  {mn@eprint@\@tempb}{\@tempb:\@tempc}{\expandafter \expandafter \csname
  mn@eprint@\@tempb\endcsname \expandafter{\@tempc}}}

\bibitem[\protect\citeauthoryear{{Althaus}, {Serenelli}, {Panei},
  {C{\'o}rsico}, {Garc{\'\i}a-Berro}  \& {Sc{\'o}ccola}}{{Althaus}
  et~al.}{2005}]{althaus05}
{Althaus} L.~G.,  {Serenelli} A.~M.,  {Panei} J.~A.,  {C{\'o}rsico} A.~H.,
  {Garc{\'\i}a-Berro} E.,   {Sc{\'o}ccola} C.~G.,  2005, \mn@doi [\aap]
  {10.1051/0004-6361:20041965}, \href
  {https://ui.adsabs.harvard.edu/abs/2005A&A...435..631A} {435, 631}

\bibitem[\protect\citeauthoryear{{Althaus}, {C{\'o}rsico}  \& {De
  Ger{\'o}nimo}}{{Althaus} et~al.}{2020}]{althaus20}
{Althaus} L.~G.,  {C{\'o}rsico} A.~H.,   {De Ger{\'o}nimo} F.,  2020, \mn@doi
  [\aap] {10.1051/0004-6361/202039557}, \href
  {https://ui.adsabs.harvard.edu/abs/2020A&A...644A..55A} {644, A55}

\bibitem[\protect\citeauthoryear{{Althaus} et~al.,}{{Althaus}
  et~al.}{2021}]{althaus21}
{Althaus} L.~G.,  et~al., 2021, \mn@doi [\aap] {10.1051/0004-6361/202038930},
  \href {https://ui.adsabs.harvard.edu/abs/2021A&A...646A..30A} {646, A30}

\bibitem[\protect\citeauthoryear{{Bailer-Jones}, {Rybizki}, {Fouesneau},
  {Demleitner}  \& {Andrae}}{{Bailer-Jones} et~al.}{2021}]{bailer21}
{Bailer-Jones} C.~A.~L.,  {Rybizki} J.,  {Fouesneau} M.,  {Demleitner} M.,
  {Andrae} R.,  2021, \mn@doi [\aj] {10.3847/1538-3881/abd806}, \href
  {https://ui.adsabs.harvard.edu/abs/2021AJ....161..147B} {161, 147}

\bibitem[\protect\citeauthoryear{{B{\'e}dard}, {Bergeron}  \&
  {Fontaine}}{{B{\'e}dard} et~al.}{2017}]{bedard17}
{B{\'e}dard} A.,  {Bergeron} P.,   {Fontaine} G.,  2017, \mn@doi [\apj]
  {10.3847/1538-4357/aa8bb6}, \href
  {https://ui.adsabs.harvard.edu/abs/2017ApJ...848...11B} {848, 11}

\bibitem[\protect\citeauthoryear{{Brassard}, {Fontaine}, {Wesemael}  \&
  {Hansen}}{{Brassard} et~al.}{1992}]{brassard92}
{Brassard} P.,  {Fontaine} G.,  {Wesemael} F.,   {Hansen} C.~J.,  1992, \mn@doi
  [\apjs] {10.1086/191668}, \href
  {https://ui.adsabs.harvard.edu/abs/1992ApJS...80..369B} {80, 369}

\bibitem[\protect\citeauthoryear{{Caiazzo} et~al.,}{{Caiazzo}
  et~al.}{2021}]{caiazzo21}
{Caiazzo} I.,  et~al., 2021, \mn@doi [\nat] {10.1038/s41586-021-03615-y}, \href
  {https://ui.adsabs.harvard.edu/abs/2021Natur.595...39C} {595, 39}

\bibitem[\protect\citeauthoryear{{Camisassa} et~al.,}{{Camisassa}
  et~al.}{2019}]{camisassa19}
{Camisassa} M.~E.,  et~al., 2019, \mn@doi [\aap] {10.1051/0004-6361/201833822},
  \href {https://ui.adsabs.harvard.edu/abs/2019A&A...625A..87C} {625, A87}

\bibitem[\protect\citeauthoryear{{C{\'o}rsico}, {Althaus}, {Montgomery},
  {Garc{\'\i}a-Berro}  \& {Isern}}{{C{\'o}rsico} et~al.}{2005}]{corsico05}
{C{\'o}rsico} A.~H.,  {Althaus} L.~G.,  {Montgomery} M.~H.,
  {Garc{\'\i}a-Berro} E.,   {Isern} J.,  2005, \mn@doi [\aap]
  {10.1051/0004-6361:20041101}, \href
  {https://ui.adsabs.harvard.edu/abs/2005A&A...429..277C} {429, 277}

\bibitem[\protect\citeauthoryear{{C{\'o}rsico}, {Althaus}, {Miller Bertolami}
  \& {Kepler}}{{C{\'o}rsico} et~al.}{2019a}]{corsico19b}
{C{\'o}rsico} A.~H.,  {Althaus} L.~G.,  {Miller Bertolami} M.~M.,   {Kepler}
  S.~O.,  2019a, \mn@doi [\aapr] {10.1007/s00159-019-0118-4}, \href
  {https://ui.adsabs.harvard.edu/abs/2019A&ARv..27....7C} {27, 7}

\bibitem[\protect\citeauthoryear{{C{\'o}rsico}, {De Ger{\'o}nimo}, {Camisassa}
  \& {Althaus}}{{C{\'o}rsico} et~al.}{2019b}]{corsico19}
{C{\'o}rsico} A.~H.,  {De Ger{\'o}nimo} F.~C.,  {Camisassa} M.~E.,   {Althaus}
  L.~G.,  2019b, \mn@doi [\aap] {10.1051/0004-6361/201936698}, \href
  {https://ui.adsabs.harvard.edu/abs/2019A&A...632A.119C} {632, A119}

\bibitem[\protect\citeauthoryear{{C{\'o}rsico}, {Althaus}, {Gil-Pons}  \&
  {Torres}}{{C{\'o}rsico} et~al.}{2021}]{corsico21}
{C{\'o}rsico} A.~H.,  {Althaus} L.~G.,  {Gil-Pons} P.,   {Torres} S.,  2021,
  \mn@doi [\aap] {10.1051/0004-6361/202040001}, \href
  {https://ui.adsabs.harvard.edu/abs/2021A&A...646A..60C} {646, A60}

\bibitem[\protect\citeauthoryear{{Curd}, {Gianninas}, {Bell}, {Kilic},
  {Romero}, {Allende Prieto}, {Winget}  \& {Winget}}{{Curd}
  et~al.}{2017}]{curd17}
{Curd} B.,  {Gianninas} A.,  {Bell} K.~J.,  {Kilic} M.,  {Romero} A.~D.,
  {Allende Prieto} C.,  {Winget} D.~E.,   {Winget} K.~I.,  2017, \mn@doi
  [\mnras] {10.1093/mnras/stx320}, \href
  {https://ui.adsabs.harvard.edu/abs/2017MNRAS.468..239C} {468, 239}

\bibitem[\protect\citeauthoryear{{De Ger{\'o}nimo}, {C{\'o}rsico}, {Althaus},
  {Wachlin}  \& {Camisassa}}{{De Ger{\'o}nimo} et~al.}{2019}]{degeronimo19}
{De Ger{\'o}nimo} F.~C.,  {C{\'o}rsico} A.~H.,  {Althaus} L.~G.,  {Wachlin}
  F.~C.,   {Camisassa} M.~E.,  2019, \mn@doi [\aap]
  {10.1051/0004-6361/201833789}, \href
  {https://ui.adsabs.harvard.edu/abs/2019A&A...621A.100D} {621, A100}

\bibitem[\protect\citeauthoryear{{Dufour}, {Blouin}, {Coutu},
  {Fortin-Archambault}, {Thibeault}, {Bergeron}  \& {Fontaine}}{{Dufour}
  et~al.}{2017}]{dufour17}
{Dufour} P.,  {Blouin} S.,  {Coutu} S.,  {Fortin-Archambault} M.,  {Thibeault}
  C.,  {Bergeron} P.,   {Fontaine} G.,  2017, in {Tremblay} P.~E.,  {Gaensicke}
  B.,   {Marsh} T.,  eds,  Astronomical Society of the Pacific Conference
  Series Vol. 509, 20th European White Dwarf Workshop. p.~3 (\mn@eprint {arXiv}
  {1610.00986})

\bibitem[\protect\citeauthoryear{{Fontaine} \& {Brassard}}{{Fontaine} \&
  {Brassard}}{2008}]{fontaine08}
{Fontaine} G.,  {Brassard} P.,  2008, \mn@doi [\pasp] {10.1086/592788}, \href
  {https://ui.adsabs.harvard.edu/abs/2008PASP..120.1043F} {120, 1043}

\bibitem[\protect\citeauthoryear{{Fontaine}, {Brassard}  \&
  {Bergeron}}{{Fontaine} et~al.}{2001}]{fontaine01}
{Fontaine} G.,  {Brassard} P.,   {Bergeron} P.,  2001, \mn@doi [\pasp]
  {10.1086/319535}, \href
  {https://ui.adsabs.harvard.edu/abs/2001PASP..113..409F} {113, 409}

\bibitem[\protect\citeauthoryear{{Giammichele} et~al.,}{{Giammichele}
  et~al.}{2018}]{giammichele18}
{Giammichele} N.,  et~al., 2018, \mn@doi [\nat] {10.1038/nature25136}, \href
  {https://ui.adsabs.harvard.edu/abs/2018Natur.554...73G} {554, 73}

\bibitem[\protect\citeauthoryear{{Gianninas}, {Bergeron}  \&
  {Ruiz}}{{Gianninas} et~al.}{2011}]{gianninas11}
{Gianninas} A.,  {Bergeron} P.,   {Ruiz} M.~T.,  2011, \mn@doi [\apj]
  {10.1088/0004-637X/743/2/138}, \href
  {https://ui.adsabs.harvard.edu/abs/2011ApJ...743..138G} {743, 138}

\bibitem[\protect\citeauthoryear{{Hermes}, {Kepler}, {Castanheira},
  {Gianninas}, {Winget}, {Montgomery}, {Brown}  \& {Harrold}}{{Hermes}
  et~al.}{2013}]{hermes13}
{Hermes} J.~J.,  {Kepler} S.~O.,  {Castanheira} B.~G.,  {Gianninas} A.,
  {Winget} D.~E.,  {Montgomery} M.~H.,  {Brown} W.~R.,   {Harrold} S.~T.,
  2013, \mn@doi [\apjl] {10.1088/2041-8205/771/1/L2}, \href
  {https://ui.adsabs.harvard.edu/abs/2013ApJ...771L...2H} {771, L2}

\bibitem[\protect\citeauthoryear{{Kanaan}, {Kepler}, {Giovannini}  \&
  {Diaz}}{{Kanaan} et~al.}{1992}]{kanaan92}
{Kanaan} A.,  {Kepler} S.~O.,  {Giovannini} O.,   {Diaz} M.,  1992, \mn@doi
  [\apjl] {10.1086/186379}, \href
  {https://ui.adsabs.harvard.edu/abs/1992ApJ...390L..89K} {390, L89}

\bibitem[\protect\citeauthoryear{{Kepler}, {Koester}  \& {Ourique}}{{Kepler}
  et~al.}{2016}]{kepler16b}
{Kepler} S.~O.,  {Koester} D.,   {Ourique} G.,  2016, \mn@doi [Science]
  {10.1126/science.aad6705}, \href
  {https://ui.adsabs.harvard.edu/abs/2016Sci...352...67K} {352, 67}

\bibitem[\protect\citeauthoryear{{Kilic}, {Bergeron}, {Kosakowski}, {Brown},
  {Ag{\"u}eros}  \& {Blouin}}{{Kilic} et~al.}{2020}]{kilic20}
{Kilic} M.,  {Bergeron} P.,  {Kosakowski} A.,  {Brown} W.~R.,  {Ag{\"u}eros}
  M.~A.,   {Blouin} S.,  2020, \mn@doi [\apj] {10.3847/1538-4357/ab9b8d}, \href
  {https://ui.adsabs.harvard.edu/abs/2020ApJ...898...84K} {898, 84}

\bibitem[\protect\citeauthoryear{{Kilic}, {Bergeron}, {Blouin}  \&
  {B{\'e}dard}}{{Kilic} et~al.}{2021a}]{kilic21}
{Kilic} M.,  {Bergeron} P.,  {Blouin} S.,   {B{\'e}dard} A.,  2021a, \mn@doi
  [\mnras] {10.1093/mnras/stab767}, \href
  {https://ui.adsabs.harvard.edu/abs/2021MNRAS.503.5397K} {503, 5397}

\bibitem[\protect\citeauthoryear{{Kilic}, {Kosakowski}, {Moss}, {Bergeron}  \&
  {Conly}}{{Kilic} et~al.}{2021b}]{kilic21b}
{Kilic} M.,  {Kosakowski} A.,  {Moss} A.~G.,  {Bergeron} P.,   {Conly} A.~A.,
  2021b, \mn@doi [\apjl] {10.3847/2041-8213/ac3b60}, \href
  {https://ui.adsabs.harvard.edu/abs/2021ApJ...923L...6K} {923, L6}

\bibitem[\protect\citeauthoryear{{Kilic} et~al.,}{{Kilic}
  et~al.}{2023}]{kilic23}
{Kilic} M.,  et~al., 2023, \mn@doi [\mnras] {10.1093/mnras/stac3182}, \href
  {https://ui.adsabs.harvard.edu/abs/2023MNRAS.518.2341K} {518, 2341}

\bibitem[\protect\citeauthoryear{{Lenz} \& {Breger}}{{Lenz} \&
  {Breger}}{2014}]{period04}
{Lenz} P.,  {Breger} M.,  2014, {Period04: Statistical analysis of large
  astronomical time series} (\mn@eprint {ascl} {1407.009})

\bibitem[\protect\citeauthoryear{{Metcalfe}, {Montgomery}  \&
  {Kanaan}}{{Metcalfe} et~al.}{2004}]{metcalfe04}
{Metcalfe} T.~S.,  {Montgomery} M.~H.,   {Kanaan} A.,  2004, \mn@doi [\apjl]
  {10.1086/420884}, \href
  {https://ui.adsabs.harvard.edu/abs/2004ApJ...605L.133M} {605, L133}

\bibitem[\protect\citeauthoryear{{Montgomery}, {Klumpe}, {Winget}  \&
  {Wood}}{{Montgomery} et~al.}{1999}]{montgomery99}
{Montgomery} M.~H.,  {Klumpe} E.~W.,  {Winget} D.~E.,   {Wood} M.~A.,  1999,
  \mn@doi [\apj] {10.1086/307871}, \href
  {https://ui.adsabs.harvard.edu/abs/1999ApJ...525..482M} {525, 482}

\bibitem[\protect\citeauthoryear{{Mukadam}, {Owen}, {Mannery}, {MacDonald},
  {Williams}, {Stauffer}  \& {Miller}}{{Mukadam} et~al.}{2011}]{mukadam11}
{Mukadam} A.~S.,  {Owen} R.,  {Mannery} E.,  {MacDonald} N.,  {Williams} B.,
  {Stauffer} F.,   {Miller} C.,  2011, \mn@doi [\pasp] {10.1086/663725}, \href
  {https://ui.adsabs.harvard.edu/abs/2011PASP..123.1423M} {123, 1423}

\bibitem[\protect\citeauthoryear{{Murai}, {Sugimoto}, {H{\={o}}shi}  \&
  {Hayashi}}{{Murai} et~al.}{1968}]{murai68}
{Murai} T.,  {Sugimoto} D.,  {H{\={o}}shi} R.,   {Hayashi} C.,  1968, \mn@doi
  [Progress of Theoretical Physics] {10.1143/PTP.39.619}, \href
  {https://ui.adsabs.harvard.edu/abs/1968PThPh..39..619M} {39, 619}

\bibitem[\protect\citeauthoryear{{Pshirkov} et~al.,}{{Pshirkov}
  et~al.}{2020}]{pshirkov20}
{Pshirkov} M.~S.,  et~al., 2020, \mn@doi [\mnras] {10.1093/mnrasl/slaa149},
  \href {https://ui.adsabs.harvard.edu/abs/2020MNRAS.499L..21P} {499, L21}

\bibitem[\protect\citeauthoryear{{Rowan}, {Tucker}, {Shappee}  \&
  {Hermes}}{{Rowan} et~al.}{2019}]{rowan19}
{Rowan} D.~M.,  {Tucker} M.~A.,  {Shappee} B.~J.,   {Hermes} J.~J.,  2019,
  \mn@doi [\mnras] {10.1093/mnras/stz1116}, \href
  {https://ui.adsabs.harvard.edu/abs/2019MNRAS.486.4574R} {486, 4574}

\bibitem[\protect\citeauthoryear{{Schwab}}{{Schwab}}{2021}]{schwab21}
{Schwab} J.,  2021, \mn@doi [\apj] {10.3847/1538-4357/abc87e}, \href
  {https://ui.adsabs.harvard.edu/abs/2021ApJ...906...53S} {906, 53}

\bibitem[\protect\citeauthoryear{{Vincent}, {Bergeron}  \&
  {Lafreni{\`e}re}}{{Vincent} et~al.}{2020}]{vincent20}
{Vincent} O.,  {Bergeron} P.,   {Lafreni{\`e}re} D.,  2020, \mn@doi [\aj]
  {10.3847/1538-3881/abbe20}, \href
  {https://ui.adsabs.harvard.edu/abs/2020AJ....160..252V} {160, 252}

\bibitem[\protect\citeauthoryear{{Winget} \& {Kepler}}{{Winget} \&
  {Kepler}}{2008}]{winget08}
{Winget} D.~E.,  {Kepler} S.~O.,  2008, \mn@doi [\araa]
  {10.1146/annurev.astro.46.060407.145250}, \href
  {https://ui.adsabs.harvard.edu/abs/2008ARA&A..46..157W} {46, 157}

\makeatother
\end{thebibliography}
\input{ms.bbl}

\bsp
\label{lastpage}

\end{document}